# First principles analysis of precipitation in Mg-Zn alloys

S. Liu[a], G. Esteban-Manzanares[a], J. LLorca[a,b,*]

[a] IMDEA Materials Institute, C/Eric Kandel 2, Getafe 28906 – Madrid, Spain
[b] Department of Materials Science. Polytechnic University of Madrid. E. T. S. de Ingenieros de Caminos. 28040 – Madrid, Spain

**Abstract**

Precipitation in Mg-Zn alloys was analyzed by means of first principles calculations. Formation energies of symmetrically distinct hcp $Mg_{1-x}Zn_x$ (0 < x < 1) configurations were determined and potential candidates for Guinier-Preston zones coherent with the matrix were identified from the convex hull. The most likely structures were ranked depending on the interface energy along the basal plane. In addition, the formation energy and vibrational entropic contributions of several phases reported experimentally ($Mg_4Zn_7$, $MgZn_2$ cubic, $MgZn_2$ hexagonal, $Mg_{21}Zn_{25}$ and $Mg_2Zn_{11}$) were calculated. The formation energies of $Mg_4Zn_7$, $MgZn_2$ cubic, and $MgZn_2$ hexagonal Laves phases were very close because they were formed by different arrangements of rhombohedral and hexagonal lattice units. It was concluded that $\beta_1'$ precipitates were formed by a mixture of all of them. Nevertheless, the differences in the geometrical arrangements led to variations in the entropic energy contributions which determined the high temperature stability. It was found that the $MgZn_2$ hexagonal Laves phase is the most stable phase at high temperature and, thus, $\beta_1'$ precipitates tend to transform into the $\beta_2'$ ($MgZn_2$ hexagonal) precipitates with higher aging temperature or longer aging times. Finally, the equilibrium $\beta$ phase ($Mg_{21}Zn_{25}$) was found to be a long-range order that precipitates the last one on account of the kinetic processes necessary to trigger the transformation from a short-range order phase $\beta_2'$ to $\beta$.

---

[*] Corresponding Author.
Email address: javier.llorca@imdea.org (J. LLorca)

# I. INTRODUCTION

Mg alloys are currently considered for structural implants in biomedical applications (bone fixation and bone scaffolds, coronary stents, etc.) due to their biocompatibility and biodegradability. They dissolve after providing support for tissue growth and/or organ healing, and it is not necessary to perform secondary surgical operation to remove the implant [1-2]. Moreover, the elastic modulus of Mg is similar to that of cortical bone avoiding problems associated with stress shielding and Mg alloys tend to promote the growth on new bone tissue [3-4]. Further developments in this area require to improve the strength and corrosion resistance of Mg alloys through the addition of alloying elements which maintain the biocompatibility and biodegradability and new alloys based on the Mg-Zn system fulfil this criterion [5].

The addition of Zn to Mg is known to lead to a remarkable age hardening response and, thus, a number of commercial alloys have been developed over the years based on the Mg-Zn binary system [6]. The age hardening results from the nucleation and growth of different precipitates from the supersaturated solid solution (SSSS) during thermal treatments. The accepted precipitation sequence is [7-14]:

$$\text{SSSS} \rightarrow \text{Guinier Preston (GP) zones} \rightarrow \beta_1' \rightarrow \beta_2' \rightarrow \beta. \tag{1}$$

The GP zones were reported to be coherent planar precipitates on the {0001} planes of α-Mg matrix [13, 15-17] formed by a single Zn atomic layer [13]. More recent investigations have shown that their thickness can reach a few Zn atomic layers [18] but their crystal structure and Zn content has not been reported [19].

$\beta_1'$ precipitates have a rodlike shape and grow parallel to the $[0001]_{\alpha\text{-Mg}}$ 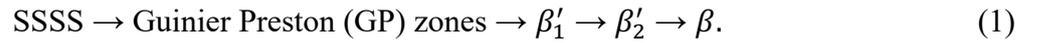 crystallographic direction during high temperature ageing. Basal slip of <a> dislocations is the softest plastic deformation mechanisms in Mg and the presence of $\beta_1'$ precipitates increases remarkably the strength of Mg alloys [20-22]. Nevertheless, the actual crystal structure of $\beta_1'$ is not well defined and it was reported in a recent study [23] that $\beta_1'$ precipitates contain a mixture of monoclinic $Mg_4Zn_7$, cubic $MgZn_2$ Laves phase and hexagonal $MgZn_2$ Laves phase. The reason why the $\beta_1'$ precipitates have such complicated structure is still unknown but this information is important to understand the details of the interactions of dislocations with these precipitates [24].

$\beta_2'$ are disc-shaped precipitates that lie on $(0001)_{\alpha\text{-Mg}}$ planes and have the structure of the $MgZn_2$ hexagonal Laves phase [6,20]. According to the experimental Mg-Zn phase diagram [25], $\beta_2'$ ($MgZn_2$) and $\beta_1'$ ($Mg_4Zn_7$) are equilibrium phases and, from a thermodynamic perspective, $\beta_1'$ precipitates should form prior to $\beta_2'$ precipitates but they seem to precipitate simultaneously in some circumstances [23].

Finally, the equilibrium $\beta$ phase was first proposed as MgZn [26], but it was later substituted by $Mg_{12}Zn_{13}$ [27] and, more recently, by $Mg_{21}Zn_{25}$ [28]. Currently, the widely accepted structure for the $\beta$ phase is $Mg_{21}Zn_{25}$ and, according to experimental phase diagram, it should precipitate before $\beta_1'$ and $\beta_2'$ but it always appears at the end of the precipitation sequence.

The analyses of the precipitate structure and precipitate sequence presented above were based on tedious experiments, involving detailed X-ray diffraction and transmission electron microscopy observations of the precipitate structure and orientation. Nevertheless, the accuracy of the results may be compromised because it is not always easy to ensure that they are representative for the whole system or a just local observations for very particular conditions. In fact, the combination of experimental observations with first principles calculations is becoming a powerful strategy to ascertain the phase stability and the precipitation sequence in metallic alloys [29-34]. In the particular case of Mg alloys, Wang et al. [31] determined the formation energy of the known stables phases in the Mg-Zn system by density functional theory (DFT) and concluded that $Mg_{21}Zn_{25}$ was the equilibrium $β$ phase. Moreover, they also reported that the formation energies of the monoclinic $Mg_4Zn_7$ and hexagonal $MgZn_2$ Laves phases were very similar, in agreement with the experimental evidence that both phases co-exist in $β_1'$ precipitates [23].

In this investigation, first principles calculations are used in combination with the cluster expansion formalism to investigate the formation energy and phase stability of intermetallic compounds in the Mg-Zn phase diagram, including the effect of the vibrational entropic contribution to the free energy at finite temperatures. Special attention was given to stablish the best candidate structures for the GP zones in the Mg-rich part of the phase diagram from the viewpoint of the formation energy. The likelihood of nucleation of each of these phases was assessed by calculating the interface energy along the coherent $(0001)_{GP}/(0001)_{Mg}$ interfaces. The results obtained by first-principles calculations indicated the most likely candidates for GP zones, rationalized the experimental evidence of the precipitation sequence in Mg-Zn alloys, and provided critical information about the stability of the different precipitates to optimize thermal treatments.

## II. METHODOLOGY

### A. First principles calculation of formation energies

Pure Mg and Zn exhibit hexagonal-close packed (hcp) structures in the stable configuration. Thus, symmetrically distinct hcp $Mg_{1-x}Zn_x$ configurations were generated with different configurations of Mg and Zn atoms in a hcp lattice sites with up to 12 atoms per unit cell using CASM (Clusters Approach to Statistical Mechanics) code [35]. The relaxed energy of each configuration was calculated by DFT using Quantum Espresso [36, 37] in the ultra-soft pseudopotential mode [38]. The atomic positions, lattice parameters and angles of each configuration were fully relaxed at pressure P=0. The exchange-correlation energy was evaluated using the Perdew-Burke-Ernzerhof approach [39] with a cut-off energy of 76 Ry. The Brillouin zone was sampled using a Monkhorst-Pack grid with a density of 30 points/Å$^{-1}$. In addition, the relaxed energies of $MgZn_2$ cubic Laves phase (cubic, Fd3m) [40], $MgZn_2$ hexagonal Laves phase (hexagonal, P6$_3$/mmc) [41], monoclinic $Mg_4Zn_7$ (monoclinic, C/2m) [42], $Mg_{21}Zn_{25}$ (trigonal, R3c) [28] and $Mg_2Zn_{11}$ (cubic, Pm3) [43] were determined by DFT using the same parameters since they are known to be the stable phases in the experimental phase diagram [25]. The evaluation of the different atomic structures was carried out using VESTA visualization code [44].

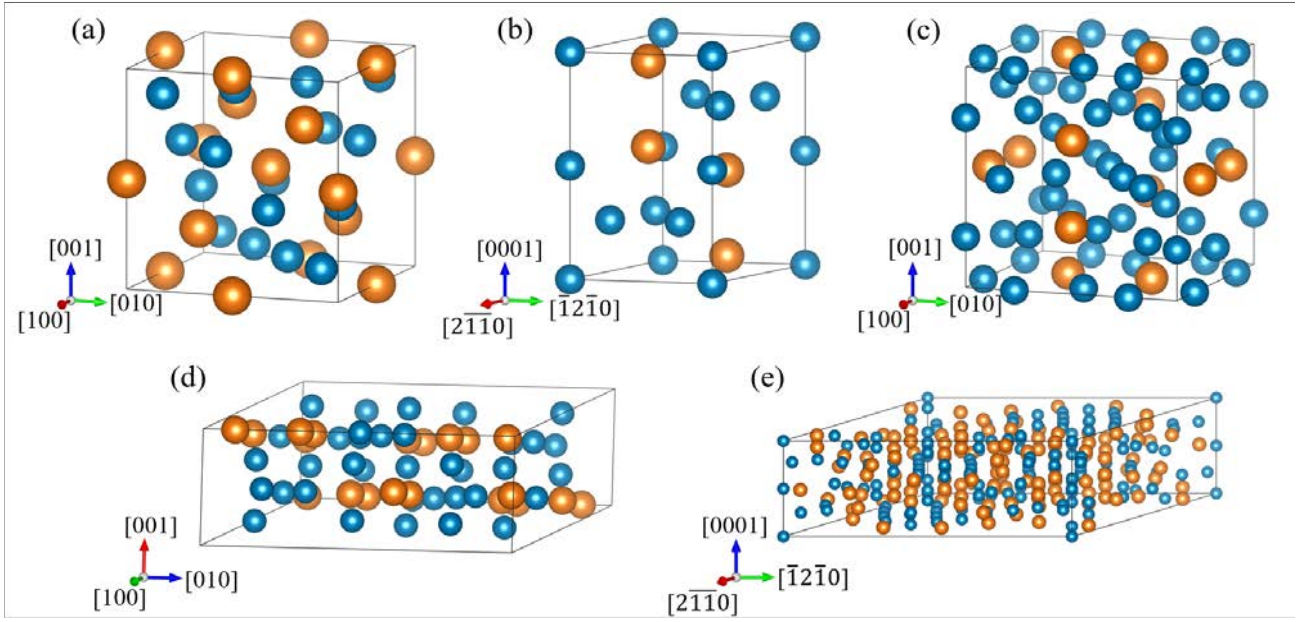

**Fig. 1.** Structure of the different intermetallic phases in the Mg-Zn phase diagram. (a) MgZn$_2$ cubic Laves phase [40]; (b) MgZn$_2$ hexagonal Laves phase [41]; (c) Mg$_4$Zn$_7$ [42]; (d) Mg$_2$Zn$_{11}$ [43]; (e) Mg$_{21}$Zn$_{25}$ [28].

The formation energy per atom of each hcp Mg$_{1-x}$Zn$_x$ configuration and each phase can be expressed as

$$E^f(Mg_{1-x}Zn_x) = E(Mg_{1-x}Zn_x) - (1-x)E(Mg) - xE(Zn) \quad (2)$$

where $E(Mg)$ and $E(Zn)$ stand for the relaxed energies per atom of pure Mg and pure Zn, and $E(Mg_{1-x}Zn_x)$ is the relaxed energy per atom of the corresponding configuration or phase.

### B. Determination of the Gibbs free energy of α-Mg

The main stable phase in the Mg-rich region of the Mg-Zn phase diagram is hcp α-Mg [25]. The equilibrium thermodynamic properties of this phase can be determined through the combination of the CE formalism with statistical mechanics principles [45-47]. The configuration ($\vec{\sigma}$) of a given crystal lattice with $N$ sites can be expressed as a collection of discrete configurational variables, e.g. $\vec{\sigma} = (\sigma_1, \sigma_2, \dots \sigma_N)$, where $\sigma_i = +1$ if the site $i$ is occupied by Mg or $\sigma_i = -1$ otherwise (Zn) in the case of a binary alloy. It has been shown that the formation energy of any configuration can be expressed as [47]

$$E^f(\vec{\sigma}) = \sum_f V_f \prod_{i \in f} \sigma_i \quad (3)$$

where $\prod_{i \in f} \sigma_i$ is a specific set of crystal basis functions that describe the different types of atomic cluster interactions (pairs, triplets, quadruplets, etc.) in the system and $V_f$ stands for the effective cluster interaction (ECI) coefficients that can be determined from the configuration energies calculated by DFT. Only a few crystal basis functions are usually necessary to obtain very accurate predictions of the formation energies, leading to an extremely efficient strategy from the computational viewpoint. It should be noted that the formation energy calculated with the CE formalism depends on the configuration and the crystal lattice and all the atomic structures considered to estimate the ECI must have the same crystal lattice. Therefore, the configurations that lost hcp crystal structure after relaxation were excluded of the process to determine the ECI coefficients, and the CE strategy in this paper only refers to the configurations that maintain the hcp structure.

In the case of the hcp Mg-Zn system, clusters of atoms with maximum atomic spacing of 10 Å for pair clusters, 8 Å for triplet clusters, and 6 Å for quadruplet clusters were considered to determine the ECI coefficients, leading to a total of 126 clusters. The optimum set of clusters and the corresponding ECIs were determined using a genetic algorithm detailed in [48] and implemented in CASM [35]. The strategy is based on the *k*-folds method were the data are divided in *k* sets (in this case *k* = 30). The ECI coefficients are calculated using *k*-1 sets as training data and the resulting values are validated on the remaining set, and this process is repeated successively for all the possible combinations of *k*-1 sets to reach the optimum ECI coefficients. The final estimator of the accuracy was a cross-validation score based on a weighted least-squares fit to obtain better predictions for the structures with lower energies. Thus, a weight is given to each configuration based on the distance $d(\vec{\sigma}) = E_{hull}^f - E^f$ to the convex hull according to $w(\vec{\sigma}) = A \exp(d(\vec{\sigma})/k_b T) + B$, where A=9, B=1 and $k_b T = 0.005$ eV. It is worth noting that the negative sign is implicit in the distance definition since all the formations energies ($E^f$) are negative and above the convex hull formation energy ($E_{hull}^f$).

The equilibrium thermodynamic properties of the α-Mg phase were obtained from the partition function *Z* using the semi-grand canonical ensemble which is given by [49]

$$Z = \sum_s e^{-\beta(E_s^f - \Delta\mu x)N} \quad (4)$$

where *N* is the number of sites in the crystal lattice, $\beta = 1/k_b T$ the Boltzmann constant and T the absolute temperature. $\Delta\mu = \mu_{Zn} - \mu_{Mg}$ is the difference of chemical potentials between Mg and Zn, *x* is the composition -expressed as the atomic fraction of Zn in the system- and $E_s^f$ stands for the formation energy of each one of the possible states of the system. Eq. (4) can be evaluated using the Metropolis Monte Carlo method [50], as detailed below, taking advantage that $E_s^f$ can be accurately and efficiently determined through eq. (3).

Once the partition function *Z* has been determined for a given temperature *T* and chemical potential $\Delta\mu$, the thermodynamic grand potential $\Phi$ can be determined as [49]:

$$\beta\Phi = -\ln Z \quad (5)$$

where

$$\Phi = U - TS - \Delta\mu x \quad (6)$$

where *U* and *S* stand for the internal energy and configurational entropy of the system.

From eqs. (4) and (5), the grand potential $\Phi$ for each phase is given by

$$d(\beta\Phi) = N(E_s^f - \Delta\mu x)d\beta - N\beta x \, d\Delta\mu \quad (7)$$

and $\Phi$ can be obtained as

$$\beta^{end}\Phi(\beta^{end}, \Delta\mu) = \beta^{begin}\Phi(\beta^{begin}, \Delta\mu) + N \int_{\beta^{begin}}^{\beta^{end}} \langle E_s^f - \Delta\mu x \rangle d\beta \quad (8)$$

for a given $\Delta\mu$ and as

$$\Phi(\beta, \Delta\mu^{end}) = \Phi(\beta, \Delta\mu^{begin}) - N \int_{\Delta\mu^{begin}}^{\Delta\mu^{end}} \langle x \rangle d\Delta\mu \quad (9)$$

for a given $\beta$. $\langle E_s^f - \Delta\mu x \rangle$ and $\langle x \rangle$ stand for the ensemble averages and $\beta^{begin}$, $\beta^{end}$, $\Delta\mu^{begin}$ and $\Delta\mu^{end}$ stand for the ranges of T and $\Delta\mu$ explored in the Metropolis Monte Carlo simulations [50].

The preliminary Monte Carlo calculation of the hcp Mg-Zn system was carried out at low temperature. A 'single-spin flip' low temperature expansion was run at 10K over a large enough range of $\Delta\mu$ (-1 eV/atom to 1 eV/atom) to obtain the grand potential references for each phase [49, 51]. Then, fine-grid metropolis Monte Carlo calculations were carried out with increasing and decreasing temperatures applying increments of 10K over the range of 10K ≤ T ≤ 700K for each chemical potential. Afterwards, the metropolis Monte Carlo calculations were performed with increasing and decreasing the chemical potentials applying increments of 0.005 eV over the range of -1 eV ≤ $\Delta\mu$ ≤ 1 eV for each temperature. The grand potential references at 10K obtained before were used as starting points, and the grand potentials for all the phases at each temperature were calculated from eqs. (8) and (9). The Monte Carlo calculations were performed in periodic supercells of 8×8×8 primitive unit cells. Equilibration of the system was carried out at all the temperatures and chemical potentials considered during this work. To this end, equilibration passes were included in every Monte Carlo simulation. Every pass included $N_{sites}$ attempted flips, with $N_{sites}$ being the number of sites in the Monte Carlo cell with variable occupations. The system was considered balanced when the precision of the sampled properties reached 95%. Afterwards, 1000 passes were carried out to calculate the thermodynamic averages. Once the grand potential has been determined, the Gibbs free energy for the ground state phases can be obtained from eq. (6) as

$$G = \Phi + \Delta\mu x \qquad (10)$$

taking into account that the differences between the Gibbs free energy and the Helmholtz free energy can be neglected in the case of solid-state transformations at atmospheric pressure because the changes in the specific volume are quite small.

## C. Vibrational contributions

The stability of the different phases in the Mg-Zn phase diagram at finite temperatures depends on the contribution of the vibrational excitacións to the formation energy and of the entropy which has two main sources in the case of metallic compounds, namely configurational (included from the partition function in eq. (5)) and vibrational entropy. Thus, the vibrational contribution to the free energy $F_v$ (which is equivalent to the Gibbs free energy the case of solid state transformations) can be expressed under the assumption that the volume of the crystal does not change with temperature as [52]:

$$F_v = E_v - TS_v(T) \qquad (11)$$

where $E_v = \int_0^\infty \frac{1}{2}\hbar\omega g(\omega)d\omega$ stands for the contribution of the vibrational excitations to the formation energy, ℏ the reduced Planck's constant, $\omega$ the volume dependent phonon frequencies and $g(\omega)$ the phonon density of states. $S_v(T)$ is the entropy associated with lattice vibration. This term can be determined as a function of temperature by means of the quasi-harmonic approximation according to [52]:

$$S_v(T) = k_b \int_0^\infty \frac{\frac{\hbar\omega}{k_bT}}{\exp\left(\frac{\hbar\omega}{k_bT}\right)-1} g(\omega)d\omega - k_b \int_0^\infty g(\omega)\ln[1-\exp\left(\frac{\hbar\omega}{k_bT}\right)]d\omega \qquad (12)$$

$g(\omega)$ was determined by the supercell method, which is based on the calculation of the forces on the atoms after perturbing slightly the atomic positions by using single point energy calculations [53]. Phonopy [54] was used to generate the perturbations in the atomic positions for each cell. 1 perturbation was generated in Mg using a 3×3×2 supercell, 1 perturbation in Zn using a 4×4×2 supercell, 12 perturbations in $MgZn_2$ hexagonal Laves phase using a 2×2×1 supercell, 18 perturbations in MgZn₂ hexagonal Laves phase using a 2×2×2 supercell and 101 perturbations in Mg₄Zn₇ using a 1×1×1 supercell. After the perturbation procedure, the constant force matrix was calculated via single point calculations and then $g(\omega)$ was estimated.

In case of line compounds with stoichiometry $Mg_{1-x}Zn_x$, whose free energy only depends on the lattice vibration (because the composition is constant), the contribution of vibrational entropy to the formation energy, $E_v^f$, was included according to

$$E_v^f(Mg_{1-x}Zn_x) = E_v(Mg_{1-x}Zn_x) - (1-x)E_v(Mg) - xE_v(Zn) \tag{13}$$

### D. Interface energy

The interface energies of the possible GP zones with respect to the α-Mg the basal plane were determined from DFT supercell calculations [55]. To this end, different supercells with the same number of $(0001)_{GP}$ and $(0001)_{Mg}$ layers containing the coherent interface $(0001)_{GP}//(0001)_{Mg}$ were constructed. The initial lattice constant *a* parallel to the interface was the average value of $a_{GP}$ and $a_{Mg}$ while the lattice constant *c* perpendicular to the interface was the one of the bulk for each compound. The relaxed energy of each supercell was calculated by DFT using Quantum Espresso and the atomic positions, lattice parameters and angles of each configuration were fully relaxed at pressure P=0. The density of k points in the Brillouin zone in these calculations was equivalent to the one used for the energy formation calculations in section II-A.

The relaxed energy of the supercell $E_{sc}$ can be expressed as [55]

$$E_{sc} = E_{Mg} + E_{GP} + \gamma\, 2A + E_{el}V \tag{14}$$

where $E_{Mg}$ and $E_{GP}$ stand for the relaxed energies of bulk Mg and GP zone, $\gamma$ is the interface energy associated with chemical bonding at the interface, $2A$ is the interface area (two interfaces appear because of the periodic boundary conditions in the DFT calculation), $E_{el}$ the elastic energy per unit volume due to the elastic mismatch between Mg and GP zone, and $V$ the volume of the supercell. Eq. (14) can be converted into:

$$\frac{\Delta E}{2A} = \frac{E_{sc} - E_{Mg} - E_{GP}}{2A} = \gamma + \frac{E_{el}V}{2A} = \gamma + \frac{E_{el}L}{2} \tag{15}$$

where L is length of the supercell perpendicular to the interface.

### III. RESULTS AND DISCUSSION
### A. Structure of potential GP zones in Mg-Zn alloys

The formation energies of 128 hcp configurations in the Mg-Zn system were obtained by DFT calculations. 18 of them lost hcp crystal structure after relaxation and they were discarded to determine the ECI coefficients. The remaining 110 are plotted in Fig. 2. The dashed red line in this figure indicates the hcp configurations with minimum formation energies and it is worth noting that 4 different configurations with x=0.17, x=0.25, x=0.33 and x=0.5 fall practically along the dashed straight line in the Mg-rich part. These four configurations have the lowest formation energy for a given Zn content with very similar chemical potentials, indicating that they are likely to precipitate at low temperatures. The arrangement of the Mg and Zn atoms in these configurations forms an ordered pattern in the (0001) plane of the hcp lattice, which is repeated along the [0001] axis (Fig. 3).

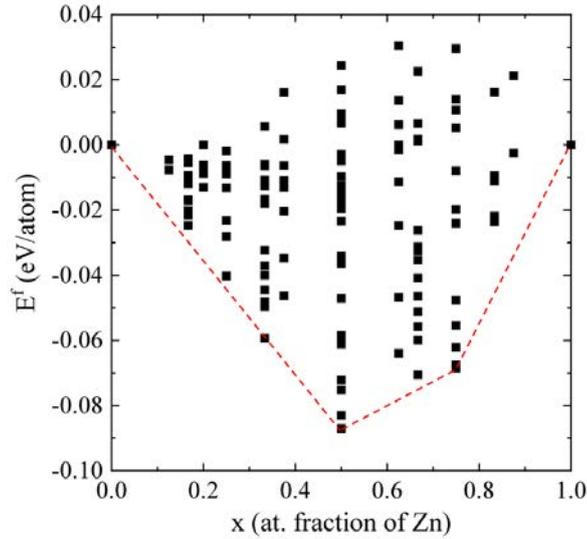

**Fig. 2.** Formation energies of hcp configurations in the Mg-Zn system calculated by DFT.

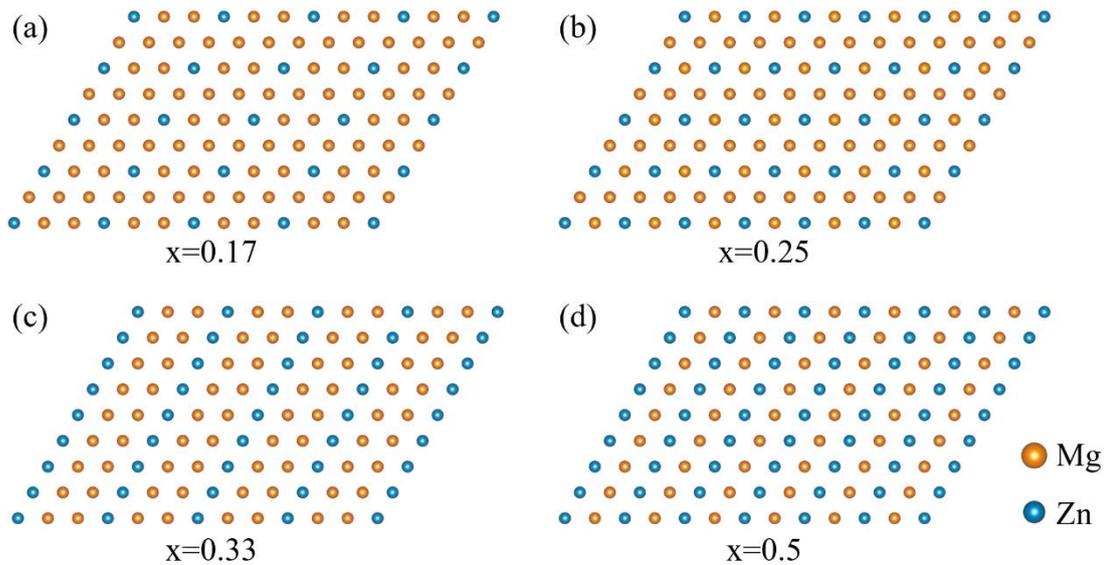

**Fig. 3.** Arrangement of Mg and Zn atoms in the basal plane of hcp lattice configurations with minimum formation energy in the Mg-rich part. $x$ stands for atomic fraction of Zn.

The ordered structures in Fig. 3 are coherent with the Mg matrix and, thus, stand for likely candidates for the GP zones in Mg-Zn alloys. Nucleation and growth of these zones from the supersaturated solid solution is driven by the minimization of the different energy contributions, namely chemical free energy, interface energy and elastic strain energy due to the lattice mismatch [52, 56]. The first one is very similar for the four potential GP zones in Fig. 3 because of the linear variation of $E^f$ with $x$ (Fig. 2) and the contribution of the interface energy (which is proportional to the precipitate surface) is normally dominant with respect to the elastic energy (which is proportional to the precipitate volume) in the case of very small coherent precipitates. Consequently, they tend to nucleate and grow along habit planes with minimum interface energy.

The interface energies of the 4 different structures in Fig. 3 along the (0001) basal plane of the hcp lattice were obtained via DFT supercell calculations. The excess of energy of the supercells (divided by the interface area) $\Delta E/2A$ is plotted in Fig. 4 as a function of the length $L$ of the supercell and the interface energy $\gamma$ for each structure is given by fitting the DFT results to eq. (15). The corresponding interface energies are depicted in Table I. They indicate that the structure with x =0.50 is the most likely candidate to form GP zones in Mg-Zn alloys followed by the one with x = 0.25.

Natarajan *et al.* [33, 57] analysed the precipitation in several Mg-RE alloys from the information provided by the formation energies at 0K calculated by DFT. Their analysis was focused in several hybrid phases labelled $\beta_P'''$ and $\beta_S'''$ in the composition range from 0 to 25 at.% RE. $\beta_P'''$ are phases containing $\beta_P'$ ordering (which has a primitive orthorhombic cell containing zig-zag rows of RE atoms along the $[2110]_{hcp}$ direction) and $\beta''$ ordering (which contains hexagon patterned RE atoms). $\beta_S'''$ are phases containing $\beta_S'$ ordering (which has a face centered orthorhombic cell containing zig-zag rows of RE atoms along the $[2110]_{hcp}$ direction) and $\beta''$ ordering. They found $\beta_P'''$ phases are more stable than $\beta_S'''$ phases in Mg-{La, Ce, Pr, Nd, Pm, Sm} alloys, while $\beta_S'''$ phases are more stable in Mg-{Sc, Y, Tb, Dy, Ho, Er, Tm, Lu} alloys. If the contribution of the misfit strains was included in the analysis, it was concluded that $\beta_P'''$ precipitates containing zig-zag and hexagon patterns will form in Mg-{La, Ce, Pr, Nd, Pm, Sm} alloys will form while $\beta_S'$ precipitates made up exclusively of zig-zag patterns will develop in in Mg-{Sc, Y, Tb, Dy, Ho, Er, Tm, Lu} alloys.

**Table I.** Interface energies (in mJ/m$^2$) of the different ordered hcp structures in Fig. 2 along the (0001) basal plane obtained from DFT calculations.

| Zn content (at. %) | Interface energy (mJ/m$^2$) |
|---|---|
| 17 | 34.3 |
| 25 | 23.2 |
| 33 | 69.7 |
| 50 | 6.0 |

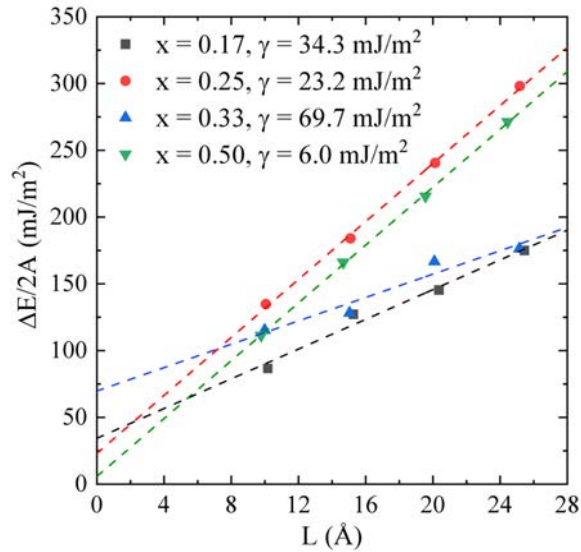

**Fig. 4.** Excess of energy of each supercell (divided by the supercell area) $\Delta E/2A$ as a function of the length L of the supercell for supercells with (0001)$_{GP}$/(0001)$_{Mg}$ interfaces for the different GP zones structures in Fig. 2.

## B. Precipitation of $\beta'_1$ and $\beta'_2$

The hcp structures in Fig. 2 are not the stable phases in the Mg-Zn system and the formation energies of the different stables phases according to the experimental evidence [6, 20-28] are plotted in Fig. 5, together with those of hcp structures in Fig. 2. The DFT calculations are in agreement with the experimental observations and indicate that $Mg_{21}Zn_{25}$, $Mg_4Zn_7$, $MgZn_2$ (cubic Laves phase), $MgZn_2$ (hexagonal Laves phase) and $Mg_2Zn_{11}$ have much lower formation energies than the corresponding hcp structures with the same composition and stand for the stable phases at low temperature. They also indicate that $Mg_{21}Zn_{25}$ is the equilibrium phase - again in agreement with experiments [28] - and that should precipitate before $Mg_4Zn_7$ and both Laves phases with $MgZn_2$ stoichiometry, which is contrary to the experimental results. This latter point will be discussed in the next section.

The results of the DFT calculations in Fig. 5 show that the formation energies of $Mg_4Zn_7$, $MgZn_2$ (cubic) and $MgZn_2$ (hexagonal) are very close. This result is in agreement with the mixture of monoclinic $Mg_4Zn_7$, cubic $MgZn_2$ and hexagonal $MgZn_2$ found in $\beta'_1$ precipitates [23] although both $Mg_4Zn_7$ and hexagonal $MgZn_2$ should be preferred to cubic $MgZn_2$ according to Fig. 5. The presence of cubic $MgZn_2$ may be explained because of a peculiar mechanism for the formation of this phase that was discovered during the analysis of the hcp lattice configurations that lost this symmetry after relaxation. In particular, the hcp lattice with alternating pure Zn (0001) layers and $\frac{2}{3}$Mg+$\frac{1}{3}$Zn (0001) layers was found to relax into the structure of the cubic $MgZn_2$ Laves phase. The shapes of the primitive cell before ($a$=$b$=5.54Å, $c$=6.04Å, $\alpha$=$\beta$=62.7°, $\theta$=60°) and after ($a$=$b$=$c$=5.21Å $\alpha$=$\beta$=$\theta$=60°) relaxation are depicted in Figs. 6(a) and (b), respectively. The equivalent atoms in both configurations are marked with the same letter. The primitive cell was only isotropic in $a$ and $b$ directions while the relaxed cell was isotropic in all directions. The Zn atoms at the vertices (marked with d) almost did not change its position, while the three Zn atoms in the cell (marked with a) and on the surface (marked with b and c) moved to the edges and became equivalent. Thus, precipitation of the $MgZn_2$ cubic Laves phase was favoured with respect to that of the $MgZn_2$ hexagonal Laves phase (although the formation energy of the latter is lower) because $MgZn_2$ cubic precipitates can be directly formed by relaxation of the ordered hcp structure in Fig. 6a without overcoming any energy barrier. As a result, simultaneous precipitation of both Laves phases has been observed in $\beta'_1$ precipitates. This type of transformation was already reported by Natarajan et al. [58]. They defined a new pathway connecting the hcp crystal structure to a hierarchy of topologically close-packed phases consisting of kagome and triangular nets (including several Laves phases).

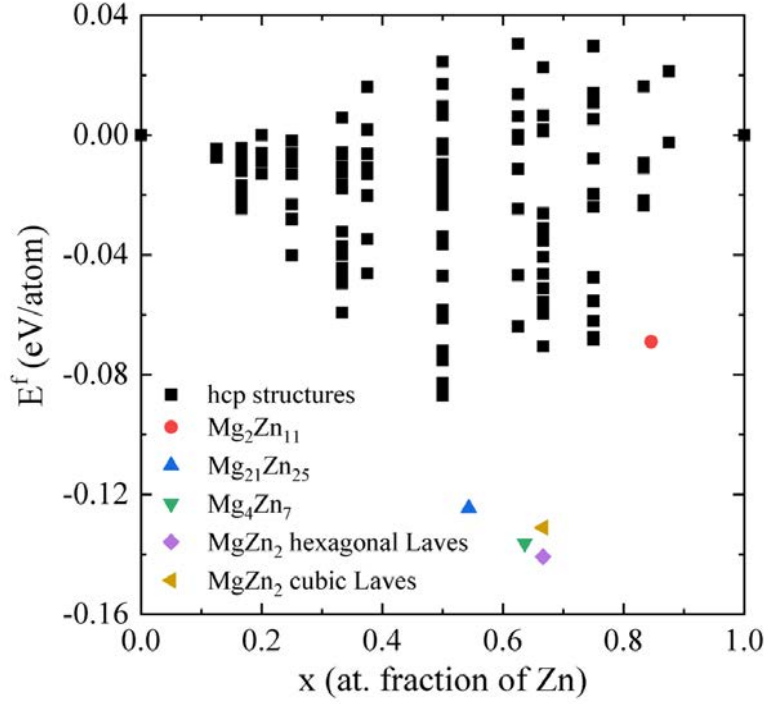

**Fig. 5.** Formation energies of $Mg_{21}Zn_{25}$, $Mg_4Zn_7$, $MgZn_2$ (cubic Laves phase), $MgZn_2$ (hexagonal Laves phase) and $Mg_2Zn_{11}$ as well as from Mg-Zn phases with hcp structure.

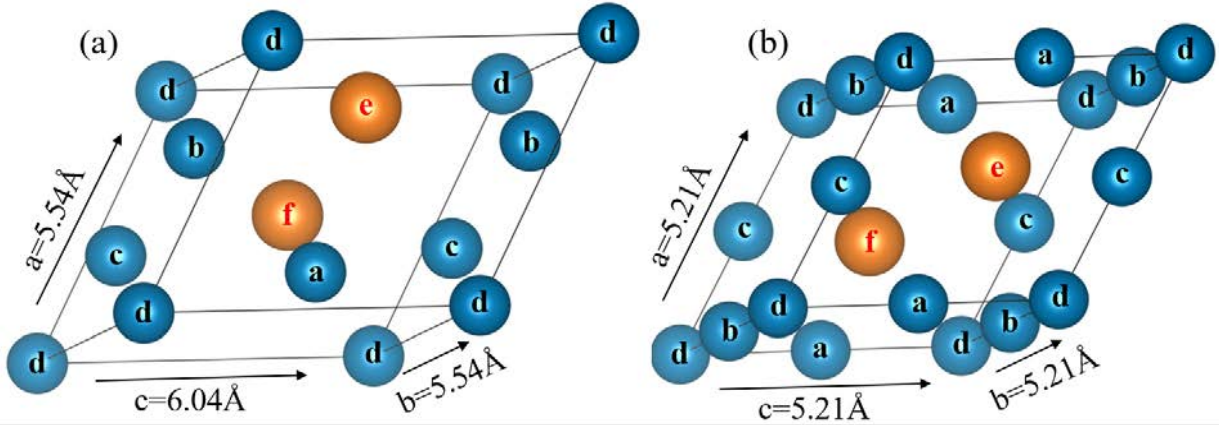

**Fig. 6.** (a) Primitive cell of a hcp lattice configuration with alternating pure Zn (0001) layers and 2/3Mg + 1/3Zn (0001) layers. (b) Cell structure after relaxation, which is the same as the primitive cell of $MgZn_2$ cubic Laves phase in Fig. 1.

It should also be noticed that high-resolution transmission electron microscopy observations of $\beta_1'$ precipitates in the $[\bar{1}\bar{1}20]$ direction of α-Mg found many similarities in the structure of these three phases [19]. In particular, $MgZn_2$ cubic and $MgZn_2$ hexagonal Laves phases can be seen as the arrangement of a series of edge-sharing rhombohedral units, each rhombohedral unit containing 2 Mg atoms and 4 Zn atoms. Nevertheless, the arrangement of the rhombohedral units is different in both $MgZn_2$ phases, as shown in Fig. 7. They are aligned in the $MgZn_2$ cubic Laves phase and form a zig-zag pattern in the $MgZn_2$ hexagonal Laves phase. Moreover, the structure of $Mg_4Zn_7$ observed from the same $[\bar{1}\bar{1}20]$ orientation contains a mixture of rhombohedral and hexagonal units, which are depicted in Fig. 7(c). The rhombohedral units are equivalent to those found in $MgZn_2$ phases and they are arranged in both aligned and zig-zag orientations. Furthermore, if the hexagonal and rhombohedral units between them are removed, the structure of the MgZn$_2$ phases is recovered. This is an interesting geometrical phenomenon associated with the isotropy of rhombohedral units in all directions and with the particular angles of the hexagonal (72º and 144º) and rhombohedral (72º and 108º)[1] units. In order to reach 360°, the angles in Mg$_4$Zn$_7$ can be assembled as 2×72°+2×108° (four neighbour rhombohedral units), 72°+72°+2×108° (three rhombohedral units and one neighbour hexagonal unit) or 144°+2×108° (two rhombohedral units and one neighbour hexagonal unit). Thus, it is not difficult to anticipate that other hexagonal and rhombohedral arrangements could lead to other phases.

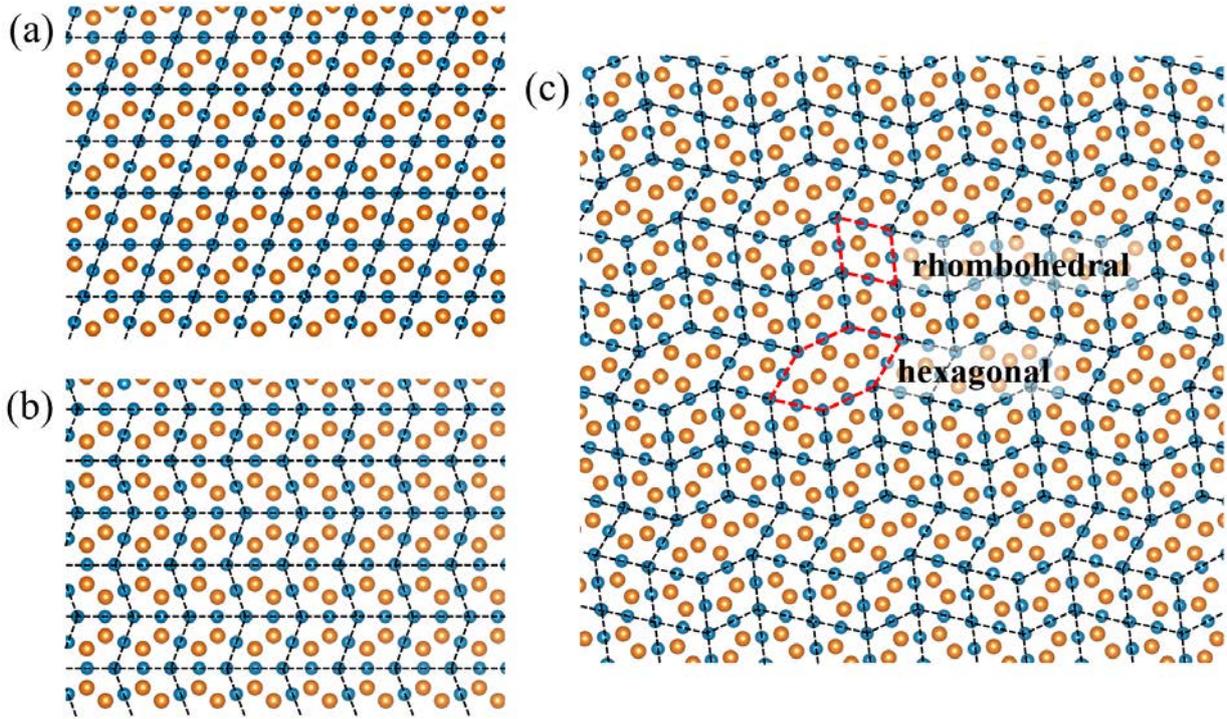

**Fig. 7.** (a) Arrangement of rhombohedral units in MgZn$_2$ cubic Laves phase. (b) Zig-zag arrangement of rhombohedral units in MgZn$_2$ hexagonal Laves phase. (c) Mg$_4$Zn$_7$ structure formed by a mixture of rhombohedral and hexagonal units. The lattice is observed from the $[\overline{1}\overline{1}20]$ orientation of α-Mg.

As mentioned above (Fig. 6), the Zn atoms at the vertices of the rhombohedral units almost did not change its position during relaxation, which mainly led to movements of the atoms in and on the unit cell surface. Therefore, if the attention is concentrated on the atoms at the vertices, it is found that the MgZn$_2$ cubic Laves phase is a fcc structure (ABCABC stacking sequence) of the rhombohedral unit while the MgZn$_2$ hexagonal Laves phase is a hcp structure (ABABAB stacking sequence) of the same unit. Therefore, the bonding environment of each atom in both Laves phases is basically identical and this similarity explains why their formation energies are so close. This is in agreement with recent investigations, which claimed that the main contribution to the chemical bonding in Laves phases come from the nearest-neighbour interactions of each atom [59-62]. The bonding environment of each atom in the hexagonal unit is also similar to that in the rhombohedral unit but with some minor changes in the number of different chemical bonds, so the formation energy of Mg$_4$Zn$_7$ is also very close. Obviously, the two Laves phases are expected to mix owing to the small energy differences and this result has also been reported for other Laves phases [63] and Mg$_4$Zn$_7$ is also expected to mix randomly with the two MgZn$_2$ Laves phases in $\beta'_1$ precipitates, as reported in

---

[1] The acute angle of the rhombohedral unit is 72°, and not 60° as it appears in Fig. 6(b), because the view direction in Fig. 7 is not perpendicular to α, β or θ.

[23]. Moreover, $\beta'_1$ precipitates are essentially a mixture of the hexagonal and rhombohedra units which lead a mixture of $Mg_4Zn_7$, $MgZn_2$ cubic Laves phase and $MgZn_2$ hexagonal Laves phase depending on the particular arrangement of the units.

It was also reported experimentally that the $\beta'_1$ precipitates tend to transform into the $\beta'_2$ precipitates, formed by the pure $MgZn_2$ hexagonal Laves phase, at higher aging temperature or longer aging times [6, 20]. In order to understand this process, the solvus curves of the $Mg_4Zn_7$, $MgZn_2$ cubic and $MgZn_2$ hexagonal phases in the hcp α-Mg phase were determined from the Gibbs free energy of α-Mg and the formation energies of the three phases including the vibrational entropic contribution.

The Gibbs free energy of the α-Mg phase was obtained from the Monte Carlo calculations detailed in section II-B with the help of the CE formalism. The formation energies of the 110 configurations used to obtain the ECIs of the CE for hcp Mg-Zn system are depicted in Fig. 2. The final optimized ECIs set includes one-point cluster interaction, one pair cluster interaction, 11 triplet cluster interactions and 3 quadruplet cluster interactions, and their values are available in Table SI in the Supplemental Material. The corresponding cross-validation score of the least-squares fitting was only 0.008 eV/atom. The Gibbs free energy of α-Mg as a function of the Zn content and temperature is depicted in Fig. 8a.

The formation energies of the different phases were calculated by including the entropic energy contribution through eqs. (11) and (13) and they are plotted as a function of temperature in Fig. 8(b). The evolution of the formation energy with temperature is very similar for both $MgZn_2$ Laves phases, which proves again the similarity both between their structures, while it grew faster for the $Mg_4Zn_7$ phase. The solvus curves of the three different phases in α-Mg were obtained by drawing the common tangent between their formation energy (Fig. 8b) and the Gibbs free energy of α-Mg (Fig. 8a). These curves are plotted in Fig. 8(c). The solvus lines of the three phases are very close at low temperatures (< 200 K) and those of $Mg_4Zn_7$ and $MgZn_2$ hexagonal phases are practically superposed up to 270K. Nevertheless, they diverge at higher temperatures and -from the viewpoint of the chemical free energy- precipitation from the supersaturated solid solution at temperatures above 350 K should begin with the $MgZn_2$ hexagonal Laves phase that is the more stable phase at high temperature. The differences in the stability between the $MgZn_2$ hexagonal and both monoclinic $Mg_4Zn_7$ and $MgZn_2$ cubic phases increase with temperature, in agreement with the experimental observation that $\beta'_1$ precipitates tend to transform into the $\beta'_2$ at high temperature or after longer aging times [6,20]. The second more stable phase is the monoclinic $Mg_4Zn_7$ up to 450K but it is replaced by the $MgZn_2$ cubic Laves phase above this temperature because of the entropic contributions to the free energy, which are plotted in Fig. 8(b). Nevertheless, it should be noted that precipitation of the $MgZn_2$ cubic phase below 450K will be favoured by the direct transformation from the ordered hcp lattice, as indicated above.

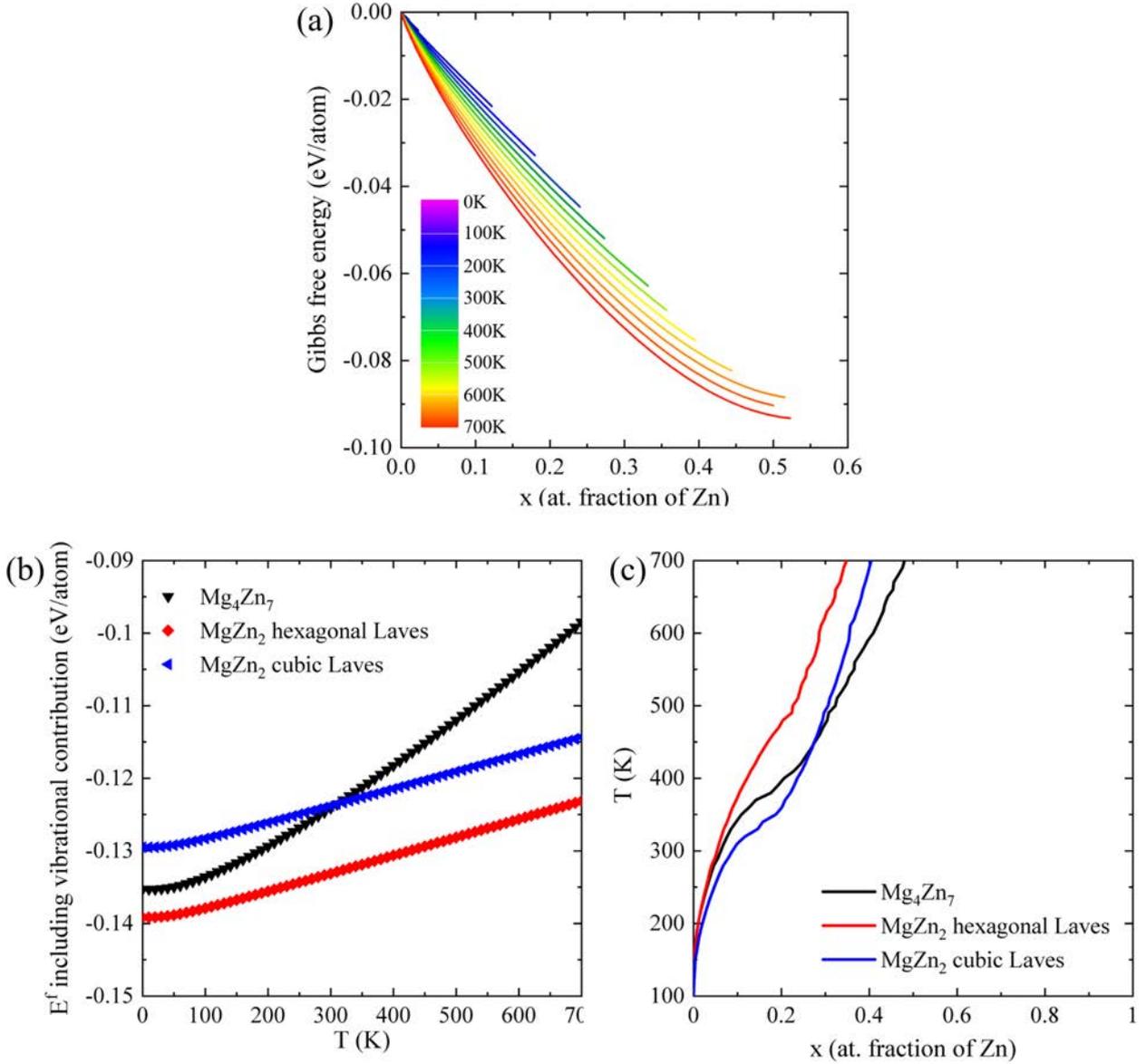

**Fig. 8.** (a) Gibbs free energies of α-Mg as a function of the Zn content and temperature. (b) Formation energies (including vibrational entropic contribution) of $Mg_4Zn_7$, $MgZn_2$ cubic Laves and $MgZn_2$ hexagonal Laves phases. (c) Solvus curves of $Mg_4Zn_7$, $MgZn_2$ cubic Laves and $MgZn_2$ hexagonal Laves phases in α-Mg.

### C. Precipitation of *β*

The formation energy of the *β* phase ($Mg_{21}Zn_{25}$) is also plotted in Fig. 5 and lies below the tie line connecting pure Mg and $Mg_4Zn_7$, indicating that is the stable phase at 0 K. Unfortunately, the unit cell of $Mg_{21}Zn_{25}$ contains 276 atoms and it is not easy to calculate the vibrational entropic contribution to assess whether this phase is still the stable phase at high temperature. Nevertheless, the experimental evidence clearly indicates that $Mg_{21}Zn_{25}$ is the equilibrium phase that appears at the end of the precipitation sequence [28] but it is not clear why this phase does not appear earlier during aging.

The structure of the $Mg_{21}Zn_{25}$ from the [0001] direction is schematically shown in Fig. 9. The unit cell is marked by rhombus. The coordinates of the atoms in the hexagons (marked by dashed red lines) differ very little from those in the $MgZn_2$ hexagonal Laves phase. The atoms outside the hexagons are at sites not far from those in $MgZn_2$ hexagonal Laves phase, but some sites occupied by Mg atoms in MgZn$_2$ are occupied by Zn atoms in Mg$_{21}$Zn$_{25}$ and vice versa. Therefore, Mg$_{21}$Zn$_{25}$ can be identified as a structure containing periodic MgZn$_2$ hexagonal Laves domains which are connected through the domain boundary, i.e. the atoms outside the hexagons. Such features are typical in long-range order structures [64] and this is in agreement with the diffraction pattern of Mg$_{21}$Zn$_{25}$ which shows superlattice diffraction peaks [65]. Thus, Mg$_{21}$Zn$_{25}$ is a long-range order phase that will precipitate after the MgZn$_2$ hexagonal Laves phase ($\beta_2'$) on account of the kinetic processes necessary to trigger the transformation from a short-range order phase (i.e. MgZn$_2$ hexagonal Laves phase) to Mg$_{21}$Zn$_{25}$. The analysis of the kinetics of precipitation of Mg$_{21}$Zn$_{25}$ is out of the scope of this paper and will be investigated in the future.

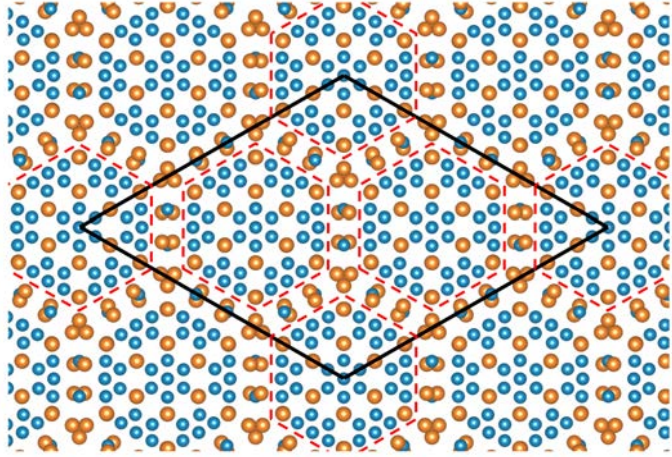

**Fig. 9.** Structure of the Mg$_{21}$Zn$_{25}$ phase observed from the [0001] direction. The unit cell is marked by a rhombus.

## IV. CONCLUSIONS

Precipitation from a supersaturated solid solution in Mg-Zn alloys was analyzed by means of first principles calculations. The formation energy of the different phases reported experimentally (Mg$_{21}$Zn$_{25}$, Mg$_4$Zn$_7$, Mg$_2$Zn$_2$ cubic Laves phase, MgZn$_2$ hexagonal Laves phase and Mg$_2$Zn$_{11}$) was calculated by DFT. They were found to have lower energies than all the hcp phases and represent the ground state phases in the Mg-Zn system. Nevertheless, several hcp structures with different Zn content (between 0.17 and 0.5 atomic fraction) and coherent with the α-Mg matrix were identified as potential candidates for the Guinier-Preston zones. Their formation energy was much smaller than the other hcp structures and all of them had very similar chemical potential. The most likely candidate was the one with the minimum (0001)$_{GP}$/(0001)$_{Mg}$ interface energy and was formed by alternating rows of Mg and Zn atoms in the basal plane.

The formation energies of monoclinic Mg$_4$Zn$_7$ and of Mg$_2$Zn$_2$ cubic and MgZn$_2$ hexagonal Laves phases obtained by DFT were very close. A careful analysis of their structures showed that they were formed by different arrangements of rhombohedral and hexagonal units in space. The bonding environment of each atom in both Laves phases is practically identical and the differences with the Mg$_4$Zn$_7$ are small, explaining the small differences in the formation energies among them. Thus, Mg$_4$Zn$_7$ is expected to mix randomly with the two MgZn$_2$ Laves phases in $\beta_1'$ precipitates, as reported experimentally. Nevertheless, the differences in the geometrical arrangement led to differences in the entropic energy contribution. It was found that the MgZn$_2$ hexagonal Laves phase is the most stable phase at high temperature and, thus, $\beta_1'$ precipitates tend to transform into the $\beta_2'$ precipitates, formed by the pure MgZn$_2$ hexagonal Laves phase, with higher aging temperature or longer aging times. Finally, the equilibrium $\beta$ phase (Mg$_{21}$Zn$_{25}$) was found to be a long-range order that precipitates the last one (even though is on the convex hull) on account of the kinetic processes necessary to trigger the transformation from a short-range order phase (i.e. MgZn$_2$ hexagonal Laves phase) to Mg$_{21}$Zn$_{25}$.

## ACKNOWLEDGEMENTS

This investigation was supported by the European Research Council (ERC) under the European Union's Horizon 2020 research and innovation programme (Advanced Grant VIRMETAL, grant agreement No. 669141). SL acknowledges the support from the European Union's Horizon 2020 research and innovation programme through a Marie Sklodowska-Curie Individual Fellowship (Grant Agreement 893883). Computer resources and technical assistance provided by the Centro de Supercomputación y Visualización de Madrid (CeSViMa) are gratefully acknowledged. Finally, use of the computational resources of the Center for Nanoscale Materials, an Office of Science user facility, supported by the U.S. Department of Energy, Office of Science, Office of Basic Energy Sciences, under Contract No. DE-AC02-06CH11357, is also gratefully acknowledged.